\documentclass[journals]{IEEEtran}
\usepackage{xspace,amsmath,amssymb,amsfonts,epsfig,subfigure,syntonly}
\usepackage{cite,bm,color,url,textcomp,empheq,boxedminipage}
\usepackage{algorithmicx,algorithm}
\usepackage{epstopdf,makecell}
\usepackage{empheq}
\usepackage{pifont}
\usepackage{subeqnarray}
\usepackage{cases}
\usepackage{graphicx,graphics}  
\usepackage{multirow,multicol}
\usepackage{psfrag}    
\usepackage{stfloats}
\usepackage{url}
\usepackage[noend]{algpseudocode}

\newtheorem{proof*}{\underline{Proof}}

\newtheorem{remark}{\underline{Remark}}[section]

\newcommand{\mv}[1]{\mbox{\boldmath{$ #1 $}}}

\psfull

\begin{document}

\title{D2D-Enabled Data Sharing for Distributed Machine Learning at Wireless Network Edge}

\author{
	\IEEEauthorblockN{Xiaoran Cai, Xiaopeng Mo, Junyang Chen, and Jie Xu}

\thanks{\scriptsize{X. Cai, X. Mo, and J. Chen are with the School of Information Engineering, Guangdong University of Technology, Guangzhou 510006, China (e-mail: xiaorancai@outlook.com, xiaopengmo@mail2.gdut.edu.cn, junyang\_chen@outlook.com). 

		J. Xu is with the Future Network of Intelligence Institute (FNii) and the School of Science and Engineering, The Chinese University of Hong Kong, Shenzhen, Shenzhen 518172, China (e-mail: xujie@cuhk.edu.cn). J. Xu is the corresponding author.}}
}

\maketitle

\begin{abstract}

	Mobile edge learning is an emerging technique that enables distributed edge devices to collaborate in training shared machine learning (ML) models by exploiting their local data samples and communication/computation resources. To deal with the “straggler’s dilemma” issue faced in this technique, this paper proposes a new device-to-device (D2D)-enabled data sharing approach, in which different edge devices share their data samples among each other over D2D communication links, in order to properly adjust their computation loads for increasing the training speed. Under this setup, we optimize the radio resource allocation for both D2D-enabled data sharing and distributed training, with the objective of minimizing the total training delay under fixed numbers of local and global iterations (for training). Numerical results show that the proposed D2D-enabled data sharing design significantly reduces the training delay, and also enhances the training accuracy when the data samples are non-independent and identically distributed (non-IID) among edge devices.

\end{abstract}

\begin{IEEEkeywords}
Mobile edge learning, data sharing, device-to-device (D2D) communications, radio resource allocation.
\end{IEEEkeywords}

\section{Introduction}

Mobile edge learning has recently attracted growing research interests from both academia and industry to enable various new artificial intelligence (AI) applications such as augmented reality (AR), industrial automation, and autonomous driving \cite{KB}. This technique aims to train machine learning (ML) models at the edge of wireless networks by exploiting data samples and communication/computation resources at distributed devices like smart phones, laptops, and smart Internet-of-things (IoT) devices. Different from the conventional ML that is normally implemented at centralized cloud, the mobile edge learning is operated at distributed edge devices and thus can efficiently reduce the traffic loads in communications networks by avoiding the long-distance data transmission from devices to cloud \cite{FL}.

The practical implementation of mobile edge learning, however, faces various technical challenges. First, the ML-training tasks are generally computation- and communication-heavy, while the edge devices are normally with small size and limited computation/communication power. Therefore, the performance of mobile edge learning is fundamentally constrained by both communication and computation at these edge devices. Next, due to the heterogeneity of edge devices, the mobile edge learning faces the so-called ``straggler's dilemma" issue, i.e., the ML-model training speed is limited by the slowest edge device in computation and communication. Furthermore, the data samples at these edge devices are generally unpredictable and may be non-independent and identically distributed (non-IID), thus making the distributed training difficult to converge and limiting the training speed and accuracy \cite{Non-iid}.

In the literature, there have been various prior studies investigating how to reduce the communication overheads in distributed ML at network edge. For instance, \cite{ComGrad1, ComGrad2} proposed to compress the exchanged gradients for reducing the communication loads. \cite{S. W. ssvm} adaptively controlled the numbers of local and global iterations during the distributed ML-model training to enhance the training speed. \cite{Ada_task} presented an adaptive task allocation scheme to deal with the ``straggler's dilemma" issue. Furthermore, \cite{Aggregation} employed the so-called ``over-the-air computation" technique for increasing the speed in aggregating the parameters/gradients from the edge devices to edge server. \cite{hierarchical_FL} investigated the distributed ML in an hierarchical system consisting of edge devices, edge servers, and cloud.

Different from the above prior works focusing on the ML strategy or structure design for improving the communication performance, this letter proposes to employ the emerging communication technique, namely the device-to-device (D2D) communications (see, e.g., \cite{D2D2, D2D}), to relieve the ``straggler's dilemma" issue for improving the performance of distributed ML-model training. Recently, the D2D communications have been recognized as one key technique in fifth-generation (5G) and beyond cellular networks, in which wireless devices in close proximity can directly communicate with each other without going through cellular infrastructures such as base stations (BSs). Motivated by this, we propose a new D2D-enabled data sharing design for mobile edge learning, which allows edge devices to share their data samples over D2D communication links. By properly controlling the amounts of data samples exchanged, this design can not only adjust the computation loads at devices for enhancing the training speed, but also reshape the data distribution (if data samples at edge devices are non-IID) for enhancing the training accuracy. In particular, we aim to minimize the total delay for the ML-model training under fixed numbers of local and global iterations (for training), by optimizing the radio resource allocation for both D2D data sharing and distributed model training. Though the formulated training delay minimization problem is non-convex, we transform it into a convex form and accordingly obtain the optimal solution. Numerical results show that our proposed D2D-enabled data sharing design efficiently speeds up the ML-model training, and also improves the training accuracy in the scenario with non-IID data distribution at edge devices.

\section{ML-Model Training via Distributed Batch Gradient Descent}

\begin{figure}[htpb]
	\centering
	\includegraphics[width=1\linewidth]{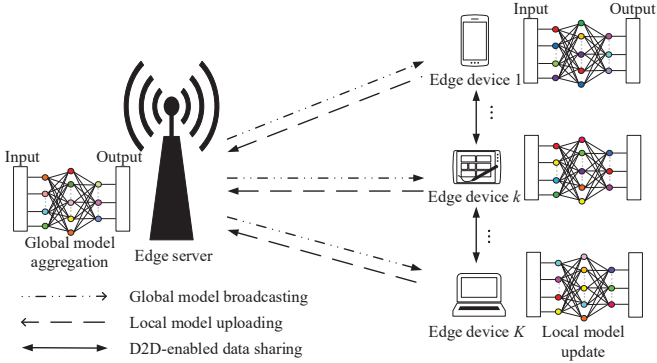}
	\caption{\small Illustration of the mobile edge learning system.}
	\label{fig1}
\end{figure}

In this section, we introduce the distributed ML-model training in a mobile edge learning system. As shown in \figurename{~\ref{fig1}}, we consider a mobile edge learning system, which consists of an edge server and a set $\mathcal{K} \triangleq \{1,\ldots,K\}$ of edge devices. Suppose that each edge device $i \in \mathcal{K}$ has a set $\mathcal{D}_i$ of training data samples, where each data sample $d \in \mathcal{D}_i$ normally contains an input vector $\boldsymbol{x}_d$ and a desired output $y_d$.\footnote{For supervised learning, the desired output corresponds to a label that is {\it a-priori} known in the training data; while for some unsupervised learning takes, the existence of $y_d$ in the training data sample may not be required \cite{S. W. ssvm}.} For each data sample, we define the corresponding loss function as $f(\boldsymbol{w}, \boldsymbol{x}_d, y_d)$, abbreviated by $f_d(\boldsymbol{w})$, where $\mv w$ denotes the ML-model parameter vector to be trained. By taking the convolutional neural network (CNN) as an example, the loss function $f_d(\boldsymbol{w})$ can be defined as the cross-entropy on cascaded linear and non-linear transforms \cite{S. W. ssvm}. Then, the local loss function at each edge device $i \in \mathcal{K}$ is given by $F_i(\boldsymbol{w}) = \frac{\sum\limits_{d \in {\mathcal{D}_i}} {{f_d}({\boldsymbol{w}})}}{|\mathcal{D}_i|}$, where $|\mathcal{A}|$ denotes the cardinality of a set $\mathcal{A}$. Accordingly, the global loss function is $F(\boldsymbol{w}) = \frac{\sum\limits_{i \in \mathcal{K}} {|\mathcal{D}_i| {F_i}({\boldsymbol{w}})}}{\sum\limits_{i \in \mathcal{K}} {|\mathcal{D}_i|}}$. The objective of mobile edge learning is to find the optimized parameter vector $\boldsymbol{w}^*$ that minimizes the global loss function $F(\boldsymbol{w})$, i.e.,
\begin{align}\label{w}
\boldsymbol{w}^* = \arg \min_{\mv w} F({\boldsymbol{w}}).
\end{align}
In order to solve problem (\ref{w}) based on data samples distributed at the $K$ edge devices, we use the distributed batch gradient descent (BGD) method, which is implemented in an iterative manner and consists of four steps at each global iteration. In the following, we explain the detailed communication and computation process in each global iteration, respectively.

\emph{I) Global Model Broadcasting:} In the first step of each global iteration, the edge server broadcasts the global parameter vector \mv w to all the $K$ edge devices for synchronization, i.e., each edge device $i \in \mathcal{K}$ updates its local model parameter vector as $\boldsymbol{w}_i = \boldsymbol{w}$. This step corresponds to the physical multicast channel in wireless communications \cite{Multicast}. Let $B$ denote the system bandwidth, $P_s$ denote the transmit power of edge server, and $g_i$ denote the channel power gain from the edge server to edge device $i \in \mathcal{K}$. Then the achievable data-rate throughput from the edge server to these edge devices is given by $r^{\text{(I)}}  = \mathop {\min }\limits_{i \in \mathcal{K}} \{ B{\log _2}(1 + \frac{{{g_i}{P_s}}}{{{n_0}B}})\}$, where $\mathit{n_0}$ denotes the power spectral density (PSD) of the additive white Gaussian noise (AWGN) at the receiver of each edge device. Furthermore, let $Q$ denote the required number of bits for sending $\mv w$, which generally depends on the quantization and compression methods used for encoding $\mv w$. Accordingly, the time duration for global model parameters downloading at each global iteration is given by
\begin{align}
t^{\text{(I)}} = \frac{Q}{\mathop {\min }\limits_{i \in \mathcal{K}} \{ B{\log _2}(1 + \frac{{{g_i}{P_s}}}{{{n_0}B}})\}}.
\end{align}

\emph{II) Local Model Update:} After synchronizing the local model parameter vector, each edge device $i \in \mathcal{K}$ updates its local model parameter vector based on the gradient of the corresponding local loss function, i.e., $\boldsymbol{w}_i \gets \boldsymbol{w}_i + \eta\nabla F_i(\boldsymbol{w}_i)$, where $\eta > 0$ denotes the learning rate, and $\nabla F_i(\boldsymbol{w}_i)$ denotes the gradient of $F_i(\boldsymbol{w}_i)$. In general, suppose that the local model update is operated over $N \ge 1$ (local) iterations to speed up the convergence by reducing the global iteration rounds. For such local model update at each edge device, we use the floating point operations (FLOPs) to measure the computation complexity, which generally depends on the considered ML-models and the size of parameter vector $\mv w$. Without loss of generality, we denote the FLOPs needed for computing the gradient for each data sample in each local iteration as $L$. Therefore, the total FLOPs required for each edge device $i$ is approximated as $NL|\mathcal{D}_i|$, and accordingly the duration of local model update at edge device $i$ is given by
\begin{align}
t^{\text{(II)}}_i = \frac{NL|\mathcal{D}_i|}{{{C_i}{f_i}}},
\end{align}
where $C_i$ denotes the FLOPs within a central processing unit (CPU) cycle and $f_i$ denotes the constant CPU frequency at each edge device $i \in \mathcal{K}$.

\emph{III) Local Model Uploading:} After all edge devices finish the local model update, they upload their local parameter vectors to the edge server. This step corresponds to a wireless multiple access channel from the $K$ edge devices to the edge server. We employ the frequency division multiple access (FDMA) transmission protocol for local model uploading, in which each edge device $i \in \mathcal{K}$ uploads its individual local model parameter vector over an orthogonal frequency band. Let $\bar b_i \ge 0$ denote the allocated (optimizable) system bandwidth for edge device $i \in \mathcal{K}$. We thus have
\begin{align}\label{bi}
\sum\limits_{i \in \mathcal{K}} {\bar b_i}  \le B.
\end{align}
Accordingly, the achievable data-rate throughput from edge device $i$ to the edge server is given by $r^{\text{(III)}}_i(\bar b_i) = \bar b_i{\log _2}\big(1 + \frac{{ {{g_{i}}} {{P_i}}}}{{{n_0}\bar b_i}}\big)$, where $P_i$ denotes the transmit power of edge device $i$. The transmission duration for local model uploading at edge device $i$ is thus given by
\begin{align}
t^{\text{(III)}}_i (\bar b_i) = \frac{Q}{\bar b_i{\log _2}\big(1 + \frac{{ {{g_{i}}} {{P_i}}}}{{{n_0}\bar b_i}}\big)}.
\end{align}

\emph{IV) Global Model Aggregation:} In the last step of the global iteration, the edge server obtains an updated parameter vector by aggregating the local models uploaded from edge devices, i.e., $\mv w \gets \frac{{\sum\limits_{i \in \mathcal{K}} {\left| {{\mathcal{D}_i}} \right|{\boldsymbol{w}_i}} }}{{\sum\limits_{i \in \mathcal{K}} {\left| {{\mathcal{D}_i}} \right|} }} $. Notice that the edge server generally has huge computation power, and as a result, the global model aggregation can be accomplished within a negligible time duration (i.e., $t^{\text{(IV)}} \approx 0$).  

As the above global iteration proceeds, the aggregated model parameter vector $\mv w$ will converge towards a desirable value. Suppose that the above processes from step \emph{I)} to \emph{IV)} are operated over $M$ (global) iterations. Then, the total duration of ML-model training (or training delay) is given by
\begin{align}\label{ML}
	t(\{\bar b_i\}) = &M\Big(t^{\text{(I)}}+ \mathop {\max } \limits_{\rm{\mathit{i} \in \mathcal{K}}} \{t^{\text{(II)}}_i + t^{\text{(III)}}_i (\bar b_i)\}\Big) \nonumber \\
	= &M\Big(\frac{Q}{\mathop {\min }\limits_{i \in \mathcal{K}} \{ B{\log _2}(1 + \frac{{{g_i}{P_s}}}{{{n_0}B}})\}} +  \nonumber \\ &\mathop {\max } \limits_{\rm{\mathit{i} \in \mathcal{K}}} \left\{ \frac{NL|\mathcal{D}_i|}{{{C_i}{f_i}}} + \frac{Q}{\bar b_i{\log _2}\big(1 + \frac{{ {{g_{i}}} {{P_i}}}}{{{n_0}\bar b_i}}\big)}\right\} \Big).
\end{align}
As the edge devices are heterogeneous in computation and communication (i.e., $|\mathcal{D}_i|, C_i, f_i, g_i, P_i$ are distinct over different edge devices), the computation and communication duration for step \emph{II)} and \emph{III)} (i.e., $t_i^{\text{(II)}} + t_i^{\text{(III)}}(\bar b_i)$) is different for different edge devices. Therefore, the slowest edge device will fundamentally limit the total duration $t(\{\bar b_i\})$. This results in the so-called ``straggler's dilemma" issue.

\section{D2D-Enabled Data Sharing with Adaptive Radio Resource Allocation} \label{III}

In this section, we propose a D2D-enabled data sharing approach to deal with the ``straggler's dilemma" issue for reducing the training delay. In particular, an additional D2D-enabled data sharing phase is implemented before the above distributed ML-model training to adjust the number of data samples (or computation loads) among the $K$ edge devices for speeding up the computation.

\subsection{D2D-Enabled Data Sharing}

In the D2D-enabled data sharing, different edge devices are enabled to exchange data samples among each other to adjust their computation loads based on their computation and communication capabilities. Let $\mathit{d_{ij}}$ $\ge 0, i, j \in \mathcal{K}, j \ne i,$ denote the number of data samples transferred from edge device $\mathit{i}$ to edge device $\mathit{j}$.\footnote{Here, $d_{ij} \in |\mathcal{D}_i|$ should be an integer. For convenience, we consider it as a non-negative real number, which is a reasonable approximation when the number of data samples at each edge device becomes large.} For each edge device $i$, we have
\begin{align}\label{dij}
\sum\limits_{j \ne i} {{d_{ij}}} \leq |\mathcal{D}_i|, \forall i \in \mathcal{K}.
\end{align}
We consider that different D2D pairs are communicated over orthogonal frequency bands. Let $b_{ij}\ge0, i, j \in \mathcal{K}, j \ne i,$ denote the bandwidth allocated in the D2D communication from edge device $i$ to edge device $j$. We have
\begin{align}\label{bij}
\sum\limits_{i \in \mathcal{K}} {\sum\limits_{j \ne i} {{b_{ij}}} }  \le B.
\end{align}
Let $p_{ij}\ge0, i, j \in \mathcal{K}, j \ne i,$ denote the transmit power at edge device $i$ for communicating with edge device $j$. Then we have
\begin{align}\label{pij}
\sum\limits_{j \ne i} {{p_{ij}}}  \le {{P_i}}, \forall i \in \mathcal{K} .
\end{align}
Consequently, the achievable D2D data-rate throughput from edge device $\mathit{i}$ to edge device $\mathit{j}$ is given by $r_{ij}({b_{ij}},{p_{ij}}) = b_{ij}{\log _2}\big(1 + \frac{{{h_{ij}}{p_{ij}}}}{{{n_0}b_{ij}}}\big)$, where $\mathit{h_{ij}}$ denotes the corresponding channel power gain. Accordingly, the transmission duration for edge device $\mathit{i}$ to transfer data to edge device $\mathit{j}$ is given as
\begin{align}
t_{ij}({d_{ij}},{b_{ij}},{p_{ij}}) = \frac{{a{d_{ij}}}}{b_{ij}{\log _2}\big(1 + \frac{{{h_{ij}}{p_{ij}}}}{{{n_0}b_{ij}}}\big)},
\end{align}
where $a$ denotes the bits of each data sample. Therefore, we obtain the total time duration for data sharing as
\begin{align}\label{dash}
	t_0({\{d_{ij}},{b_{ij}},{p_{ij}}\}) = \max_{ i, j \in \mathcal{K},\atop \scriptsize j \ne i} \left\{ \frac{{a{d_{ij}}}}{b_{ij}{\log _2}(1 + \frac{{{h_{ij}} {p_{ij}}}} {{{n_0}b_{ij}}})}\right\} .
\end{align}

After finishing data sharing, the number of data samples at edge device $i \in \mathcal{K}$ is expressed as
\begin{align}
	|\mathcal{D}_i'| = |\mathcal{D}_i| + \sum\limits_{j \ne i} {{d_{ji}} - \sum\limits_{j \ne i} {{d_{ij}}}},
\end{align}
where $\mathcal{D}_i'$ denotes the updated data set at edge device $i$. In this case, the delay for each edge device $i$ to execute the local model update (Step \emph{II)}) is revised as
\begin{align}
\bar t^{\text{(II)}}_i({\{d_{ij}\}}) = \frac{{NL({|\mathcal{D}_i| + \sum\limits_{j \ne i} {{d_{ji}} - \sum\limits_{j \ne i} {{d_{ij}}} }})}}{{{C_i}{f_i}}}.
\end{align}
Therefore, under D2D-enabled data sharing, the total time duration for the ML-model training or total training delay is given by
\begin{align} \label{t}
&\bar t({\{d_{ij}}, {b_{ij}}, {p_{ij}}, \bar b_i\}) \nonumber \\
= & t_0\big({\{d_{ij}},{b_{ij}},{p_{ij}}\}\big) +  M\big(t^{\text{(I)}} + \mathop {\max }\limits_{\rm{\mathit{i} \in \mathcal{K}}} \{\bar t ^{\text{(II)}}_i({\{d_{ij}\}}) + t^{\text{(III)}}_i (\bar b_i)\}\big) \nonumber \\
= & \max_{ i, j \in \mathcal{K},\atop \scriptsize j \ne i} \left\{ \frac{{a{d_{ij}}}}{b_{ij}{\log _2}(1 + \frac{{{h_{ij}} {p_{ij}}}} {{{n_0}b_{ij}}})}\right\} + M\Big(\frac{Q}{\mathop {\min }\limits_{i \in \mathcal{K}} \{ B{\log _2}\big(1 + \frac{{{g_i}{P_s}}}{{{n_0}B}}\big)\}} +  \nonumber \\ &\mathop {\max } \limits_{\rm{\mathit{i} \in \mathcal{K}}} \left\{ \frac{{NL({|\mathcal{D}_i| + \sum\limits_{j \ne i} {{d_{ji}} - \sum\limits_{j \ne i} {{d_{ij}}} }})}}{{{C_i}{f_i}}} + \frac{Q}{\bar b_i{\log _2}\big(1 + \frac{{ {{g_{i}}} {{P_i}}}}{{{n_0}\bar b_i}}\big)}\right\} \Big).
\end{align}
By comparing $\bar t({\{d_{ij}}, {b_{ij}}, {p_{ij}}, \bar b_i\})$ in (\ref{t}) versus $t(\{\bar b_i\})$ in (\ref{ML}), it is observed that the D2D-enabled data sharing can reshape the data samples at edge devices for reducing the computation delay at step \emph{II)} at the cost of introducing an additional delay term $t_0({\{d_{ij}},{b_{ij}},{p_{ij}}\})$. Therefore, to minimize the total training delay, we need to design the data sharing $\{d_{ij}\}$ and the radio resource allocation $\{b_{ij}, p_{ij}, \bar b_i\}$ for balancing such a trade-off.

\subsection{Adaptive Radio Resource Allocation}

In this subsection, we optimize the radio resource allocation for both D2D-enabled data sharing and distributed training to minimize the total training delay. For ease of notation, we define $\mv v$ as a vector containing all the variables in $\{d_{ij}, b_{ij}, p_{ij}, \bar b_i\}$, and we use $\mv v \succeq \mv 0$ to denote that all these variables are non-negative. Therefore, the radio resource allocation problem for minimizing the training delay is formulated as
\begin{align}
	\text{(P1)}: \min_{ \mv v \succeq \mv 0} ~&\bar t({\{d_{ij}}, {b_{ij}}, {p_{ij}}, \bar b_i\}) \label{ObjP2} \\
	\text{s.t. } &(\ref{bi}), (\ref{dij}), (\ref{bij}), (\ref{pij}).\nonumber
\end{align}
Note that problem (P1) is generally challenging to be solved as the objective function in (\ref{ObjP2}) is non-convex due to the coupling of $d_{ij}$, $b_{ij}$ and $p_{ij}$.

To solve problem (P1), we first introduce two auxiliary variables $\tau_1$ and $\tau_2$, and accordingly transform problem (P1) as the following equivalent problem:
\begin{subequations}\label{Prob2.1}
	\begin{align}
	\text{(P1.1)}: ~&\min_{\scriptsize \tau_1, \tau_2, \mv v}  \tau_1 + M(t^{\text{(I)}} + \tau_2) \label{objP2.1}\\
	\text{s.t. } &\tau_1 \geq 0, \tau_2 \geq 0 \label{ConstP2.1_1} \\
	&ad_{ij} \leq \tau_1 b_{ij}{\log _2}\big(1 + \frac{{{h_{ij}}{p_{ij}}}}{{{n_0}b_{ij}}}\big), \forall i, j \in \mathcal{K}, j \ne i \label{ConstP2.1_3} \\
	&\bar t^{\text{(II)}}_i({\{d_{ij}\}}) + t^{\text{(III)}}_i (\bar b_i) \leq \tau_2, \forall i \in \mathcal{K}, j \ne i \label{ConstP2.1_4} \\
	&(\ref{bi}), (\ref{dij}), (\ref{bij}), (\ref{pij}). \nonumber
	\end{align}
\end{subequations}
	Problem (P1.1) is still non-convex due to the coupling of $b_{ij}$ and $\tau_1$ in (\ref{ConstP2.1_3}). Nevertheless, notice that problem (P1.1) is a convex problem under any given $\tau_1$. Therefore, we propose to solve problem (P1.1) and thus (P1), by optimizing over $\mv v$ and $\tau_2$ by using CVX \cite{cvx} under any given $\tau_1$, and then using a one-dimensional (1D) search over $\tau_1$. During the 1D search, the regime of $\tau_1$ is set to be $[0, T_1]$, where $T_1$ is given in (\ref{T1}), which corresponds to the training delay under the case without data sharing and equal bandwidth allocation (i.e., $d_{ij} = 0, \forall i,j \in \mathcal{K}, i \neq j$, and $\bar b_i =\frac{B}{K}$).
\begin{align} \label{T1}
T_1 =  &M\Big(\frac{Q}{\mathop {\min }\limits_{i \in \mathcal{K}} \{ B{\log _2}\big(1 + \frac{{{g_i}{P_s}}}{{{n_0}B}}\big)\}} +  \nonumber \\ &\mathop {\max } \limits_{\rm{\mathit{i} \in \mathcal{K}}} \left\{ \frac{NL|\mathcal{D}_i|}{{{C_i}{f_i}}} + \frac{Q}{\frac{B}{K}{\log _2}\big(1 + \frac{{ {{g_{i}}} {{P_i}}}}{{{n_0}\frac{B}{K}}}\big)}\right\} \Big).
\end{align}
Therefore, problem (P1) is finally solved optimally.

\begin{remark}
	It can be shown that at the optimal solution to (P1), we must have $d_{ij}^* \cdot d_{ji}^* = 0, \forall i, j \in \mathcal{K}, j \ne i$, since otherwise, we can always modify $\{d_{ij}\}$ to achieve the same data distribution but with shorter delay of $t_0\big({\{d_{ij}},{b_{ij}},{p_{ij}}\}\big)$ for data sharing. This shows that the data sharing between any two edge devices must be unidirectional.
\end{remark}

\begin{remark}
	It can also be shown that at the optimality of (P1), it must hold that $\bar t^{\text{(II)}}_i ({\{d_{ij}^*\}}) + t^{\text{(III)}}_i (\bar b_i^*) = \bar t^{\text{(II)}}_j({\{d_{ji}^*\}}) + t^{\text{(III)}}_j (\bar b_j^*), \forall i, j \in \mathcal{K}, j \ne i$, since otherwise, we can further reduce the total training delay by adjusting the bandwidth and power allocations. This shows that under the adaptive radio resource allocation, all edge devices should finish their local model update and uploading at the same time for minimizing the overall training delay.
\end{remark}

\section{Numerical Results} 

In this section, we present numerical results to validate the performance of our proposed D2D-enabled data sharing design, as compared with the following two benchmark schemes.

\begin{itemize}
	\item {\it \textbf{Conventional design with fixed radio resource allocation:}} In this scheme, no D2D-enabled data sharing is implemented with $d_{ij} = 0, \forall i,j\in \mathcal{K}, i\neq j$. In step \emph{III)} at each global iteration, the edge devices upload their updated local ML-model parameters to the edge server under equal bandwidth allocation, i.e., $\bar b_i = \frac{B}{K}, \forall i \in \mathcal{K}$. The corresponding training delay is given by $T_1$ in (\ref{T1}).
	
	\item {\it \textbf{Conventional design with adaptive radio resource allocation:}} In this scheme, no D2D-enabled data sharing is implemented. Adaptive radio resource allocation is implemented in step \emph{III)} of each global iteration. Accordingly, the bandwidth allocation for minimizing the training delay is formulated as
	\begin{align}
	\text{(P2)}: \min_{ \{\bar b_i\}}~& t(\{\bar b_i\}) \label{ObjP1} \nonumber\\ 
	\text{s.t. } &(\ref{bi}),\nonumber
	\end{align}
	which can be transformed into a convex form similarly as for (P1) and then solved optimally via CVX.
\end{itemize}

In the simulation, we consider that there are $K=6$ edge devices, which are located at a distance of $350$ meters with the edge server. We consider the path loss model ${\beta _0}({d}/{d_0}){^{ - \alpha }}$, where $\beta_0 = -30$ dB denotes the path loss at the reference distance of $d_0 = 1$ m, $\alpha=3$ is the pathloss exponent and $d$ is the distance between the transmitter and the receiver. Unless otherwise stated, we set the transmit power of the edge server as $P_\text{s}=43$ dBm, the system bandwidth as $B=1$ MHz, and the noise PSD as $n_0 = -130$ dBm/Hz. For edge devices, we set the number of FLOPs within a CPU cycle as $C_1=C_2=8$, $C_3=C_4=12$, $C_5=C_6=16$, the CPU frequency as $f_1=f_2=1.5$ GHz, $f_3=f_4=1.95$ GHz, $f_5=f_6=2.5$ GHz, and the transmit power as $P_i=33$ dBm$, \forall i \in \mathcal{K}$.

Furthermore, we consider that a CNN\footnote{The considered CNN has 7 Layers: $5\times 5 \times 32$ convolutional layer $\rightarrow 2 \times 2$ maxPool layer $\rightarrow 5\times 5 \times 32$ convolutional layer $\rightarrow 2 \times 2$ maxPool layer $\rightarrow 1568 \times 256$ fully connected layer $\rightarrow 256 \times 10$ fully connected layer $\rightarrow$ log-softmax layer.} \cite{CNN} is trained by using the distributed BGD. We consider the MNIST dataset \cite{MNIST}, where each image has 784 pixels and a label. For the purpose of initial investigation, we quantize each pixel into 8 bits without data compression used. Furthermore, suppose that each label consists of 4 bits. Accordingly, we have $a = 784 \times 8 + 4 = 6,276$ bits for each data sample. We also quantize each element in the ML-model parameter vector into $8$ bits. It then follows from \cite{FLOPs} that under our considered CNN model, we have $Q = 3.2$ Gbits and $L \approx 6$ GFLOPs. Furthermore, we set the number of local iterations as $N = 5$ and the learning rate as $\eta = 0.01$.

In addition, we consider the non-IID data distribution. In the considered MNIST dataset, there are 10 different types of labels. Accordingly, we divide the MNIST dataset into 10 subsets, and choose 5,000 data samples for each edge device from only two different subsets.

\begin{figure}[htbp]
	\centering
	\includegraphics[width=1\linewidth]{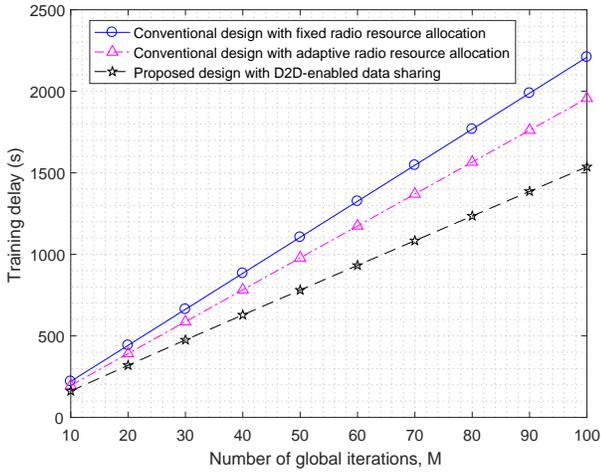}
	\caption{\small The total training delay versus the number of global iterations $M$.}
	\label{mtcnn}
\end{figure}

\figurename{~\ref{mtcnn}} shows the total time duration versus the number of global iterations $M$. It is observed that our proposed design with D2D-enabled data sharing achieves considerably shorter training delay than the two benchmark schemes. Such benefits are gained by reshaping the data distributions (and equivalently computation loads) among these devices, together with the adaptive radio resource allocation. Furthermore, such performance gain is observed to become more substantial as $M$ increases.

\begin{figure}[htbp]
	\centering
	\includegraphics[width=1\linewidth]{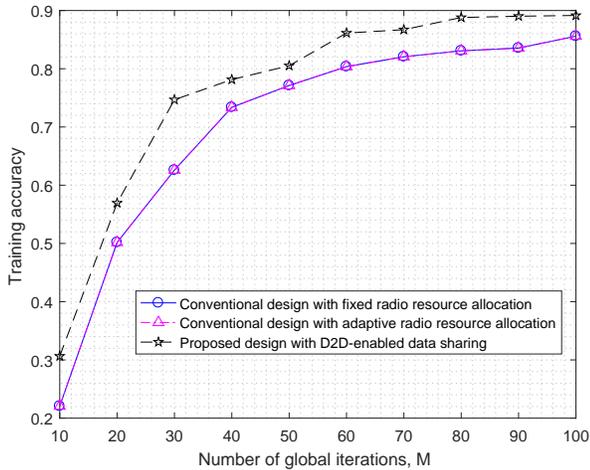}
	\caption{\small The training accuracy versus the number of global iterations $M$.}
	\label{acnn}
\end{figure}

\figurename{~\ref{acnn}} shows the training accuracy versus the number of global iterations $M$. It is observed that our proposed design with D2D-enabled data sharing achieves much higher training accuracy than the two benchmark schemes. This validates that the D2D-enabled data sharing is also beneficial in changing the data distribution among edge devices to resolve the non-IID distribution issue, thus improving the training accuracy.

\section{Conclusion}

This letter proposed a D2D-enabled data sharing design for mobile edge learning, in which different edge devices can exchange their data samples among each other for reshaping the computation loads and data distributions. Under this setup, we proposed to optimize the adaptive radio resource allocation to minimize the total training delay. Numerical results showed that the proposed design not only improves the training speed, but also enhances the training accuracy, as compared to conventional designs without such consideration. 


\end{document}